\documentclass[useAMS,usenatbib]{mn2e}
\usepackage{graphicx,amssymb,amsmath,times}
\title{Effect of nearby supernova remnants on local Cosmic-Rays}
\author[Satyendra Thoudam]{Satyendra Thoudam\thanks{E-mail: satyend@barc.gov.in}\\
Astrophysical Sciences Division,\\
Bhabha Atomic Research Centre, Trombay, Mumbai-400085,\\
Maharashtra, India}
\begin{document}
\date{}
\pagerange{}
\maketitle
\label{firstpage}
\begin{abstract}
We study in detail the effect of different particle release times from sources on the cosmic-ray (CR) spectrum below $10^{15}eV$ in the Galaxy. We discuss different possible forms of particle injection such as burst-like injection, continuous injection for a finite time, injection from a stationary source and energy dependent injection. When applied to the nearby known supernova remnants, we find that the observed CR anisotropy data favour the burst-like particle injection model for the CR  diffusion coefficient $D(E)\propto E^a$ with $a=0.3-0.6$ in the local region. In the study we have also found that the contribution of the sources G114.3+0.3 and Monogem dominate if the observed anisotropy is a result of the effect of the nearby sources. Further study shows that we should not neglect the contribution of the \textit{undetected} old sources to the local CR anisotropy. 
\end{abstract}

\section{Introduction}
Considering the similarity of the power supplied by a supernova explosion (SNE), which is $\sim 10^{42}erg/s$, and the power required to maintain the cosmic-ray (CR) energy density in the Galaxy ($\sim 10^{41}erg/s$), it is widely beleived that the majority of CRs in the Galaxy are accelerated in supernova remnant (SNR) shock waves. If this is true, then it is very important to know the maximum energy of CRs that can be accelerated in SNRs as well as the fraction $f$ of SNE energy that can be converted into CRs by diffusive shock acceleration. Studies based on the plane-wave approximation have shown that the maximum energy can be $\sim Z\times10^{14} eV$,where Z is the charge of the particle (see e.g. Lagage $\&$ Cesarsky 1983). However, studies using nonlinear effects produced by accelerated CRs have found that the maximum energy can be as large as $\sim Z\times10^{15} eV$ (Berezhko 1996). Regarding the total energy transfered into CRs in a SNE, there can actually be a considerable variation in the value of $f$ among the sources, but its value is generally considered to be $\sim 0.1$ which can easily be up to $\sim 0.3$ (Berezhko et al. 1996). 

An interesting feature of the CRs observed at the Earth is that they are highly isotropic. Their anisotropy amplitude is only $\sim (10^{-4}-10^{-3})$ in the energy range of $(10^{11}-10^{15})eV$ (Guillian et al. 2007 and references therein) with the phase mainly found in the outer Galaxy particularly in the second quadrant of the Galaxy. The high degree of isotropy is because CRs do not reach Earth from the sources directly. They undergo difusion in the Galaxy as a result of scattering either by magnetic field irregularities or by self-excited Alfven and hydromagnetic waves before reaching the Earth. In the present study, because CRs are composed mainly of protons below the knee region $(\sim 3\times10^{15}eV)$, the anisotropy below the knee is considered mainly because of the proton component.     

The anisotropy observed at the Earth may be because of the effect of global diffusion leakage of CRs from the Galaxy, the effect of random nature of CR sources both in space and time, or the effect of the local sources (see e.g. Ptuskin et. al. 2005). Studies using local SNRs lying within $1kpc$ from us with ages less than $10^5 yr$ have been carried out, assuming these to be instantaneous point sources (Dorman et al. 1984). It was concluded that the fact that the CR anisotropy is almost energy independent in the energy range of $(10^{12}-10^{14})eV$ may be because we are observing the Vela source's maximum anisotropy amplitude.   

In a recent paper, Erlykin $\&$ Wolfendale (2006) have shown that the observed characteristics of CRs, and in particular their anisotropy, are mainly the product of local sources located within a distance of $(1-2) kpc$ from the Sun. Their study considers the actual positions of the observed SNRs within $3kpc$ from the Sun as well as the position of the solar system at the inner edge of the Orion arm. However, they could not incorporate the actual ages of the  SNRs in their study because their model considers the CRs to be confined in the SNRs for around $8\times10^4 yr$, but the majority of observed SNRs have ages less than $\sim 4\times10^4 yr$. Among the ten samples that they simulated, only one sample has been  found to show the phase of the anisotropy which agrees with the observed data.  

Considering local SNRs located within a distance of $1kpc$ and age less than $5\times10^4 yr$, a study had also been carried out using the data on the high energy CR spectrum and anisotropy to constrain the value of CR diffusion coefficient in the Galaxy and its dependence on the energy (Ptuskin et al. 2005), assuming the particles are released from the sources at the time of the explosion itself. Ptuskin et al. have shown that the diffusion model with reacceleration is barely compatible with the observed CR anisotropy, with a discrepancy that is roughly within a factor of 3. The plain diffusion model predicts a value of anisotropy that is too large. 

From these earlier works, it can be seen that the cause of the observed amplitude and the phase of the CR anisotropy is not clearly understood, although local phenomena are more likely to play a dominant role. In the present study we also try to understand the problem, but with more emphasis given to the particle release time. We consider different possible forms of particle injection from the sources, such as like burst-like injection, continuous injection for a finite time, injection from a stationary source and energy dependent injection. We then study the observed local SNRs and examine the effect of the different particle release times on the observed CR anisotropy. In short, the plan of the paper can be outlined as follows. In Section 2 we calculate the CR spectrum generated by a single point-like source. In Section 3 we discuss the effect of different particle release times from the sources on the CR flux and anisotropy. In Section 4 we give an application to the nearby SNRs located within $1.5 kpc$ from the Earth. Finally,  in Section 5 we present the results and discussions about the whole study.

\section {CRs from a single point like source}

\subsection{CR spectrum}
In the diffusion model, neglecting convection, energy losses and particle losses as a result of nuclear interactions, the propagation of CR protons in the Galaxy is given by (see e.g. Gaisser 1990 and references therein)
\begin{equation}
\nabla\cdot(D\nabla N)+Q=\frac{\partial N}{\partial t}
\end{equation} 
where $N(\textbf{r},E,t)$ is the differential number density, $E$ is the proton kinetic energy, $D(E)\propto E^a$ with $a=$ constant (positive) is the diffusion coefficient which is assumed to be spatially uniform in the Galaxy and $Q(\textbf{r},E,t)$ is the proton production rate [i.e. $Q(\textbf{r},E,t) d^3\textbf{r} dE dt$ is the number of protons produced by the source in a volume element $d^3\textbf{r}$ in the energy range $(E,E+dE)$ in time $dt$].

The Green function $G(\textbf{r},\textbf{r}^\prime,t,t^\prime)$ of Eq. (1) [i.e. the solution for a $\delta$-function source term $Q(\textbf{r},t)=\delta(\textbf{r}-\textbf{r}^\prime)\delta(t-t^\prime)$] is obtained as
\begin{equation}
G(\textbf{r},\textbf{r}^\prime,t,t^\prime)=\frac{1}{8\pi^{3/2}\left[D(t-t^\prime)\right]^{3/2}}exp\left[\frac{-(\textbf{r}^\prime -\textbf{r})^2}{4D(t-t^\prime)}\right]
\end{equation} 
Then, the general solution of Eq. (1) is given by
\begin{equation}
N(\textbf{r},E,t)=\int^{\infty}_{-\infty}d\textbf{r}^\prime\int^t_{-\infty}dt^\prime G(\textbf{r},\textbf{r}^\prime,t,t^\prime)Q(\textbf{r}^\prime,E^\prime,t^\prime)
\end{equation}
Now, to calculate the number density $N(\textbf{r},E,t)$ at a distance $\textbf{r}$ from a point source, let us consider a small source volume $\varsigma$ where protons are generated. Before the protons are released into the interstellar medium (ISM), they are assumed to be distributed uniformly in the source volume with density $q_0(E)$. In other words, we can write 
\begin{eqnarray}
Q(\textbf{r},E,t)=q_0(E)q(t),\qquad inside\;the\;source,\nonumber\\
=0,\qquad\qquad\quad\; outside\;the\;source, 
\end{eqnarray} 
and for a point source\\ 
\\
$\lim_{\varsigma\to 0}q_0(E)\varsigma=q(E)$
\\
With this assumption the proton density at a distance $\textbf{r}$ from a point source can be calculated by introducing Eqs. (2)$\&$(4) in Eq. (3) as
\begin{eqnarray}
N(\textbf{r},E,t)=\frac{1}{8\pi^{3/2}}\int^t_{-\infty}dt^\prime\frac{1}{\left[D(t-t^\prime)\right]^{3/2}}\nonumber\\
\times exp\left[\frac{-r^2}{4D(t-t^\prime)}\right]q(E)q(t^\prime)
\end{eqnarray}
and the proton flux is given by $I(E)\approx (c/4\pi)N(E)$, where $c$ is the velocity of light. The proton source spectrum $q(E)$ is taken as
\begin{equation}
q(E)=k(E^2+2Em_p)^{-(\Gamma+1)/2}(E+m_p)
\end{equation}
where $k$ is the normalization constant and $m_p$ is the proton mass energy. The source spectral index $\Gamma$ is chosen such that $\Gamma + a=2.73$, the observed proton spectral index (Haino et al. 2004). In Eq. (5), the function $q(t^\prime)$ carries the information about how and when the CR particles are emitted by the source. The actual form of $q(t^\prime)$ is not known exactly. It can be a simple energy independent function or a complicated energy dependent function also (i.e. particles with different energies emitted at different times). In fact, detailed studies have shown that the most energetic paricles start to escape from the source region at the very beginning of the Sedov phase itself, although the major fraction of the CRs remain confined for a period of about $10^5 yr$ (Berezhko et al. 1996). In the present study, we will consider different possible forms of $q(t^\prime)$'s corresponding to different types of particle injection, that is, burst-like injection, continuous injection for a finite time, injection from a stationary source and also the energy dependent injection mentioned above. 

\subsection{CR anisotropy}
The amplitude of CR anisotropy due to a single source in the diffusion approximation is given by (Mao $\&$ Shen 1972)
\begin{equation}
\delta_i=\frac{3D}{c}\frac{|\nabla N_i|}{N_i}
\end{equation}
where $N_i$ is given by Eq. (5) for a point source $i$ with distance $\textbf{r}_i$ and age $t_i$. The total anisotropy parameter at the Earth, due to a number of nearby discrete sources in the presence of an isotropic CR background, is given by 
\begin{equation}
\delta=\frac{\displaystyle\sum_{i} I_i\delta_i \hat{r}_i.\hat{n}_m}{I_T}
\end{equation} 
where the summation is over the nearby discrete sources. $\hat{r}_i$ denotes the direction of the source $i$ giving a flux $I_i$ and $\hat{r}_m$ denotes the direction of maximum intensity. $I_T=1.37(E/GeV)^{-2.73} cm^{-2}s^{-1}sr^{-1}GeV^{-1}$ represents the total observed flux of CR protons above $\sim 10GeV$ (Haino et al. 2004). The direction (phase) of the anisotropy is given by the direction of maximum intensity. Therefore, the anisotropy $\delta$ as well as the phase at an energy $E$ depends on the age and distance of the nearby sources. These can be determined by different sources at different energy intervals. In fact, the energy dependence of $\delta$ depends on the dependence of $I_i,I_T$ and $\delta_i$ on energy and are discussed in detail in the next section for different particle injection forms. In case a single source is dominant over the entire energy range, the total anisotropy $\delta$ is given by 
\begin{equation}
\delta=\frac{I_m}{I_T}\delta_m
\end{equation}
where the subscript $m$ denotes the source giving the maximum flux at the Earth.

\section{Effect of particle release time from the sources}

\subsection{Burst- like injection}
When the injection time of CRs from the sources is significantly short, the injection process can be represented by a $\delta$-function as
\begin{equation}
q(t^\prime)=\delta (t^\prime-t_0)
\end{equation}
Eq. (10) represents the burst-like injection of all the particles at time $t_0$ after the explosion. Using this, the proton density is obtained from Eq. (5) as
\begin{equation}
N(\textbf{r},E,t)=\frac{q_b(E)}{8\pi^{3/2}\left[D(t-t_0)\right]^{3/2}}exp\left[\frac{-r^2}{4D(t-t_0)}\right]
\end{equation} 
For the present case, the source spectrum $q_b(E)$ is given by Eq. (6) with the constant $k$ obtained by normalizing the proton source spectrum such that the total amount of kinetic energy contained in the injected protons is $f\sim (0.1-0.3)$ times the total SNE energy. With this, the single source anisotropy (Eq. 7) becomes
\begin{equation}
\delta_i=\frac{3}{2c}\frac{r_i}{(t_i-t_0)}
\end{equation}
which is independent of energy. For the high energy particles for which the diffusive time $t_d\sim r^2/D$ taken to travel a distance $r$ in the Galaxy is less than the time interval after their release from the source $(t-t_0)$, i.e. $t_d<(t-t_0)$, the exponential term in Eq. (11) tends to 1 and hence, the proton spectrum follows a simple power law as $N(E)\propto E^{-(\Gamma+3a/2)}$. Similar result can also be found in Aharonian $\&$ Atoyan 1996 where they studied about the pion decay gamma radiation in the vicinity of the CR sources. Also, the total proton spectrum observed at the Earth $I_T$ can be written in the form $I_T(E)\propto E^{-(\Gamma+a)}$ where $\Gamma+a=2.73$ as mentioned in section 2.1. Therefore, the total anisotropy $\delta$ (Eq. 9), when a single source dominates, is found to follow an energy dependence of the form $\delta\propto E^{-a/2}$. $\delta$ also depends strongly on the time $t_0$ after the SNE at which particles are injected into the ISM. Applying Eqs. (11)$\&$(12) into Eq. (9), the dependence of high energy particles on injection time is given by $\delta\propto(t_m-t_0)^{-5/2}$ where $t_m$ denotes the age of the source giving the maximum intensity at an energy $E$ at the Earth. This shows that as the value of $t_0$ increases, the total CR anisotropy increases.  

\subsection{Continuous injection for a finite time}
For the case of continuous injection of CRs from a source for a finite time interval from $t_1$ to $t_2$ where $(0\leq t_1<t_2)$, the source is switched on only between $t_1$ and $t_2$. So, the temporal particle injection function can be written as
\begin{eqnarray}
q(t^\prime)=1,\;\;\; for\;\;t_1\leq t^\prime\leq t_2,\nonumber\\
=0,\qquad\;\;\;\;\;\; otherwise,
\end{eqnarray}
The source spectrum (Eq. 6) for this case is given by 
\begin{equation}
q_c(E)=\frac{q_b(E)}{(t_2-t_1)}
\end{equation} 
where $q_b(E)$ has been obtained in section 3.1. The proton density can then be easily obtained from Eq. (5) as
\begin{equation}
N(\textbf{r},E,t)=\frac{q_c(E)}{4\pi rD}\left[erf(\sqrt{x_2})-erf(\sqrt{x_1})\right]
\end{equation} 
where,
\begin{equation}
x_1=\frac{r^2}{4D(t-t_1)}\;\;\; and\;\;\;x_2=\frac{r^2}{4D(t-t_f)}\;,
\end{equation}
$erf(y)=(2/\sqrt{\pi})\int_0^y exp(-u^2)du$ is the error function and $t_f=min[t,t_2]$. The corresponding single source anisotropy amplitude (Eq. 7) is given by
\begin{equation}
\delta_i=\frac{3D}{cr_i}\left[1+\frac{2}{\sqrt{\pi}}\left\lbrace\frac{\sqrt{x_1}exp(-x_1)-\sqrt{x_2}exp(-x_2)}{erf(\sqrt{x_2})-erf(\sqrt{x_1})}\right\rbrace\right]
\end{equation}
It is possible to discuss the particle spectrum given by Eq. (15) for two seperate cases: $t_1<t\leq t_2$ and $t_1<t_2\leq t$. The two cases are expected to give completely different spectra, particularly for the higher energy particles as discussed below.
\\
\\
\textit{Case 1:} $(t_1<t\leq t_2):-$ For this case, we observe the source while it is still active and liberating CRs continuously into the ISM. Here, $t_f=t$ which implies $x_2=\infty$ (using Eq. 16). By the property of error function, $erf(\sqrt{x_2})=1$ for $x_2=\infty$ (Abramowitz $\&$ Stegun 1964). So, Eq. (15) for this case can be written as
\begin{equation}
N(\textbf{r},E,t)=\frac{q_c(E)}{4\pi rD}\left[1-erf(\sqrt{x_1})\right]
\end{equation}  
For the high energy particles with a diffusion time $t_d<<(t-t_1)$, $x_1\rightarrow 0$ and because $erf(\sqrt{x_1})\rightarrow 0$ for $x_1\rightarrow 0$, the spectrum given by Eq. (18) follows a power-law of the form $N(E)\propto E^{-(\Gamma+a)}$ and the single source anisotropy (Eq. 17) has an energy dependence given by $\delta_i\propto E^a$. Thus, the total anisotropy (Eq. 9) varies with energy as $\delta\propto E^a$ in the high energy regime in this case.
\\
\\
\textit{Case 2:} $(t_1<t_2\leq t):-$ For this case, $t_f=t_2$ and hence, $x_2=r^2/4D(t-t_2)$. Here, the proton spectrum is given by Eq.(15), but with $t_f$ replaced by $t_2$. For very high energy particles for which $x_1\rightarrow 0$, we can safely write $x_2<<1$ as $t_1<t_2$. However, a property of the error function gives $erf(\sqrt{x_2})\approx 2\sqrt{x_2/\pi}$ for $\sqrt{x_2}<<1$. So, for high energy particles the spectrum can be written as
\begin{equation}
N(\textbf{r},E,t)\approx \frac{q_c(E)}{4(\pi D)^{3/2}\sqrt{t-t_2}}
\end{equation}
and the single source anisotropy can be written as
\begin{equation}
\delta_i\approx \frac{3}{4c}\frac{r_i}{(t_i-t_2)}
\end{equation} 
The energy dependence of Eqs. (19)$\&$(20) is the same as in the case of burst-like injection of particles discussed in section 3.1. This implies that the total anisotropy parameter (Eq. 9) varies as $\delta\propto E^{-a/2}$.

From the study, we can now see that for the continuous injection of particles it is important to know when the observation is being made for a particular source. If we observe the source after it is switched off, we see a spectrum much steeper than would have been observed during its active period. For the present study, we take $t_2=10^5 yr$ because it is generally accepted that CRs remain confined in the SNRs upto around $10^5 yr$ (see section 3.4). So, for the nearby SNRs which are listed in Table 1, it is expected that for the same source spectrum the Geminga supernova will give a steeper spectrum at the Earth compared to all the other remaining SNRs, because it is the only source (with an estimated age of $t\sim 3.4\times 10^5 yr$) that is being observed after it is switched off. Another important point is that for a fixed value of $t_2$, the total anisotropy parameter $\delta$ (Eq. 9) varies with $t_1$. In both cases discussed above, this is given by $\delta\propto1/(t_2-t_1)$ for high energy particles. Thus, the dependence of $\delta$ on injection time in the continuous particle injection model is weaker than that in the burst-like injection model.
  
\subsection{Injection from a stationary source}
The particle spectrum for the case of injection of CR particles from a stationary source can be easily obtained using Eq. (11). For the burst-like injection of particles at time $t_0=0$, Eq. (11) gives 
\begin{equation}
N(\textbf{r},E,t)=\frac{q_b(E)}{8\pi^{3/2}(Dt)^{3/2}}exp\left[\frac{-r^2}{4Dt}\right]
\end{equation}
For a stationary source injecting a $q_s(E)$ number of particles of energy $E$ per unit time continuously, the particle density $N(\textbf{r},E)$ is obtained by replacing $q_b(E)$ in Eq. (21) with $q_s(E)dt$ and integrating over $dt$ from $0$ to $\infty$ as
\begin{equation}
N(\textbf{r},E)=\frac{q_s(E)}{4\pi rD}
\end{equation}
This shows that the particle spectrum follows a simple power-law of the form $N(E)\propto E^{-(\Gamma+a)}$ as in the continuous injection case (Case 1 of section 3.2). The single source anisotropy amplitude in this case is
\begin{equation}
\delta_i=\frac{3D}{cr_i}
\end{equation}
and the total anisotropy (Eq. 9) follows $\delta\propto E^a$. 
 
\subsection{Energy dependent particle injection}
Studies based on diffusive shock acceleration in SNRs have shown that the highest energy particles already start leaving the source region at the beginning of the Sedov phase (Berezhko et al. 1996), but the major fraction of accelerated CRs remain confined for almost around $10^5yr$ for an ISM hydrogen atom density of $n_H=1 cm^{-3}$. Thus, an energy dependent proton confinement time in the source region (proton release time from the source) can be written as (Berezhko $\&$ V$\ddot{o}$lk 2000)
\begin{equation}
T_{con}(E)=min[10^5,10^3(E/E_{max})^{-5}]yr
\end{equation}
where $E_{max}$, the maximum energy of accelerated particles, is taken as $10^{15} eV$ (Berezhko 1996) and $10^3 yr$ marks the start of the Sedov phase. For such a type of particle injection, the function $q(t^\prime)$ in Eq. (5) can be written as
\begin{equation}
q(t^\prime)=\delta(t^\prime-T_{con})
\end{equation} 
The particle spectrum and the anisotropy parameter for this case are the same as in the burst-like injection case with $t_0$ replaced by $T_{con}(E)$. From Eq. (24), it can be seen that all the particles with energies less than $\sim 4\times 10^{14} eV$ remain confined until $10^5 yr$. So, except for particles with $E\geq 4\times 10^{14} eV$ both the spectrum and the anisotropy are the same as those of the burst-like injection of particles at time $t_0=10^5 yr$.

\section{Application to nearby SNRs}
The diffusion coefficient $D(E)\propto E^a$, which governs the diffusion of CRs in the Galaxy, is generally obtained using the observed secondary to primary boron-to-carbon (B/C) ratio. However, the values of $D(E)$ obtained from the same experimental data are different for different CR propagation models in the Galaxy. The values vary from $D(E)=2\times 10^{28}(E/5GeV)^{0.6}cm^2s^{-1}$ for $E>5 GeV$, where $E$ is in GeV (Engelmann et al. 1990), obtained using a leaky box propagation model, to $D(E)=5.9\times 10^{28}\beta E^{0.3}cm^2s^{-1}$, where $\beta=v/c$ and $v$ is the velocity of the particle (Jones et al. 2001), given by a diffusion model with particle reaccelaration. Here, we consider both cases for the present study with the source spectral index $\Gamma$ chosen such that $\Gamma +a=2.73$.    

Considering SNE at various distances from the Earth, a recent study has shown that only those sources located within a distance of $\sim 1.5 kpc$ can produce appreciable temporal fluctuations in the CR proton flux observed at the Earth (Thoudam 2006). So, for the present study we consider only those SNRs whose ages are known and which are lying within $1.5 kpc$ from us. These SNRs are listed in Table 1 along with their galactic longitudes $(l)$, distances $(d)$ and ages $(t)$. The SNR RXJ1713.7-3946 is not included in the list because of the uncertainty in its distance. Its distance of $(6\pm 1) kpc$ is in contrast to the distance of $1 kpc$ estimated from the soft X-ray obsorption (Koyama et al. 1997).

\begin{table}
\centering
\caption{Parameters of SNRs with known ages located within a distance of $1.5 kpc$ from the Earth (References: $^1$ Braun et al. 1989; $^2$ Strom 1994 and for other sources, see the references given in Thoudam 2006).}
\begin{tabular}{@{}lllrrlrlr@{}}
\hline
SNR &  $l(deg.)$ & $d(kpc)$      &$t(yr)$\\         
\hline
G65.3+5.7   & 65.3		&     1.0  &   14000\\
G73.9+0.9   & 73.9		&     1.3  &   10000\\
Cygnus Loop & 74.0		&     0.4  &   14000\\
HB21        & 89.0		&     0.8  &   19000\\
G114.3+0.3  & 114.3		&     0.7  &   41000\\
CTA1        & 119.5		&     1.4  &   24500\\
HB9         & 160.9		&     1.0  &   7700 \\
S147\footnote&180.0		&	   0.8 &   4600\\
Vela        & 263.9		&     0.3  &   11000 \\
G299.2-2.9  & 299.2		&     0.5  &   5000\\
SN185\footnote&315.4	&	   0.95&	 1800\\
Monogem     & 201.1		&     0.3  &   86000\\
Geminga     & 195.1		&     0.4  &   340000\\
\hline
\end{tabular}
\end{table}

Among the various particle injection schemes considered in section 3, only the burst-like and the continuous injections are considered for application to the nearby SNRs. The case of injection from a stationary source is not considered, mainly because every source has a finite lifetime and should also have a finite particle injection time. Moreover, the proton spectrum and the anisotropy $\delta$ follow the same energy dependence as in the case of continuous injection for a finite time period (see section 3.3). Similarly, energy dependent particle injection is the same as burst-like injection at time $t_0=10^5 yr$ except for particles with energies $E\geq4\times 10^{14} eV$ as discussed in section 3.4.
\begin{figure}
\centering
\includegraphics*[width=0.31\textwidth,angle=270,clip]{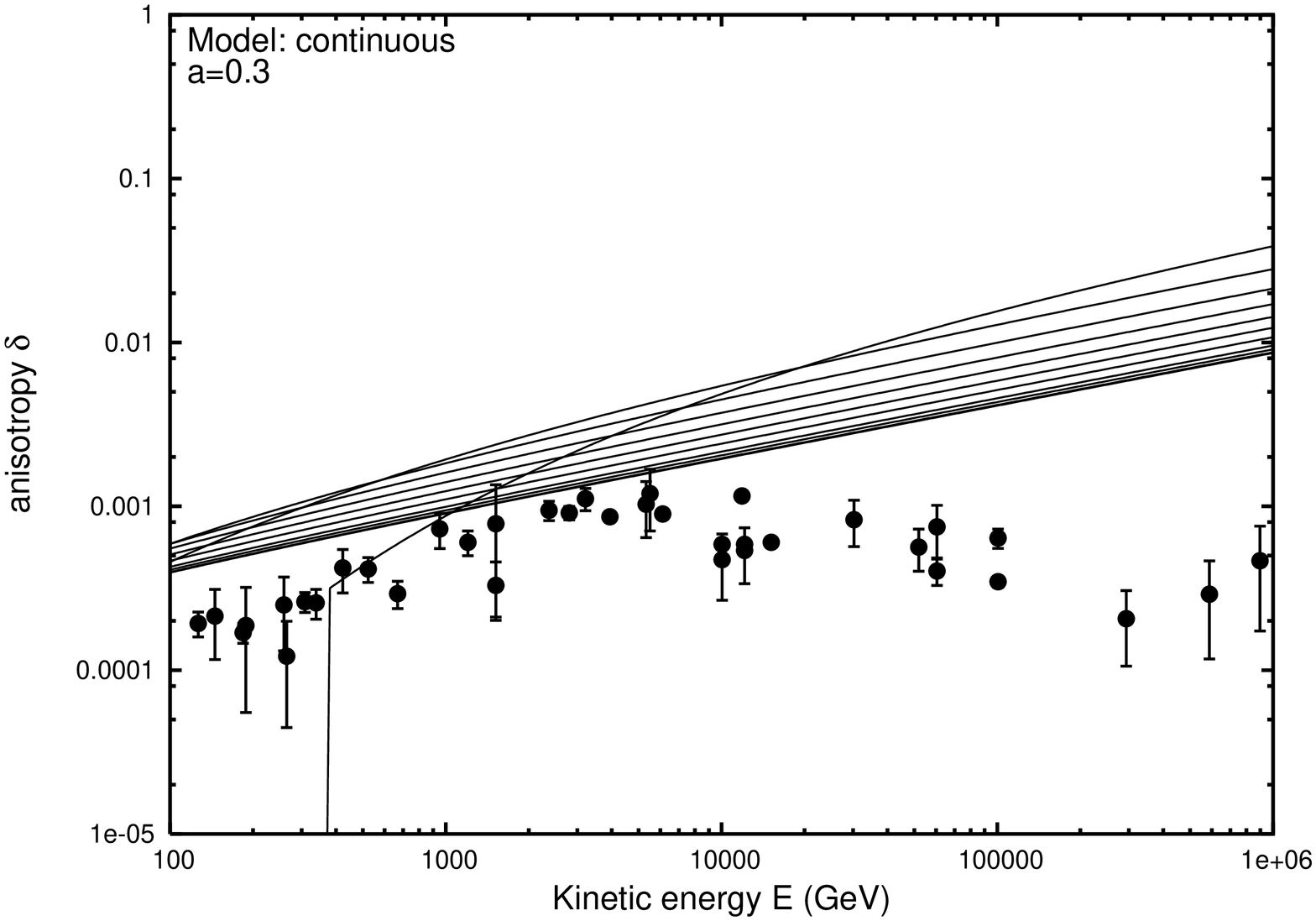}
\includegraphics*[width=0.31\textwidth,angle=270,clip]{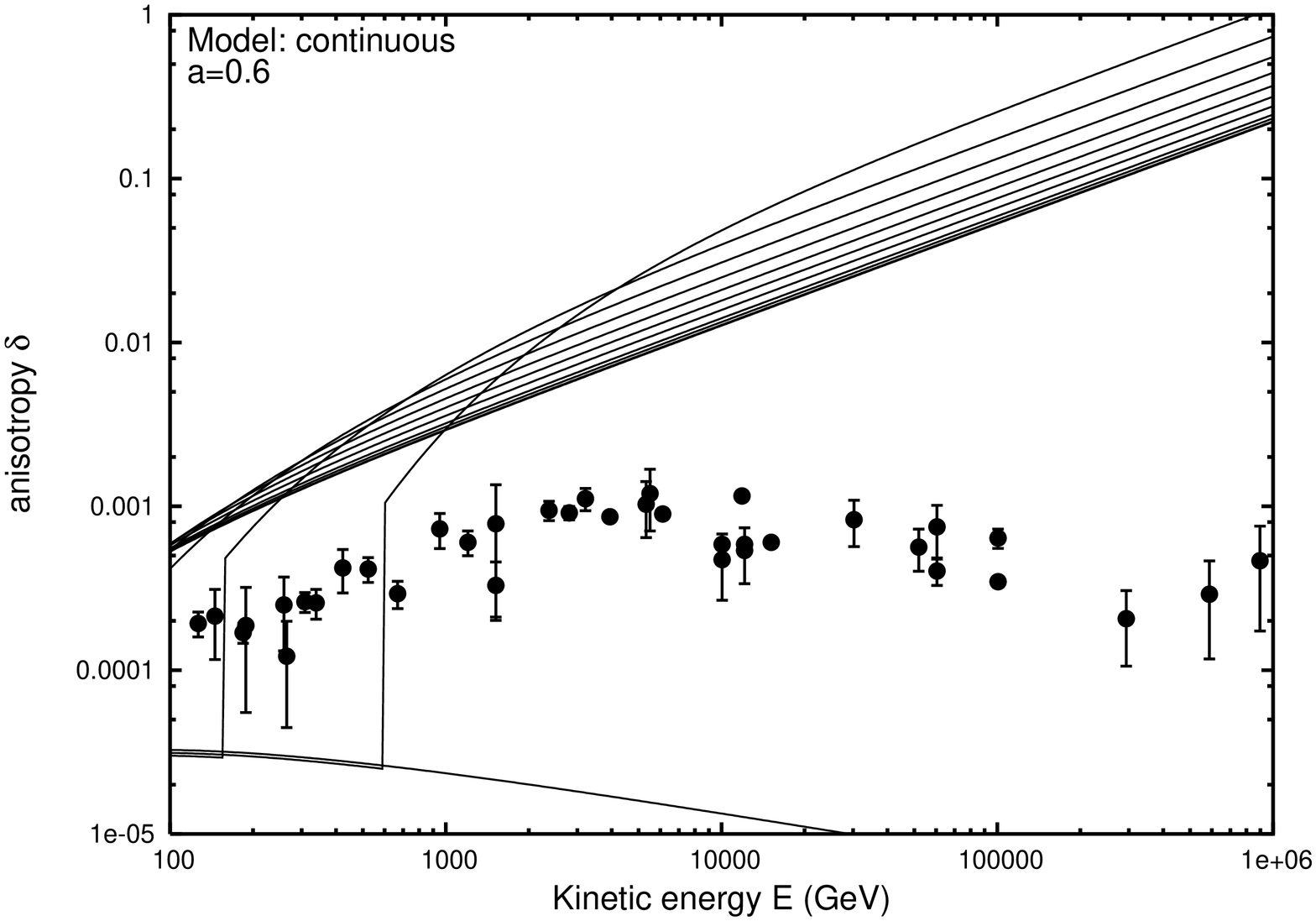}
\caption{\label {fig1} CR anisotropy $\delta$ (Eq. 8) for the continuous injection case. The solid lines represent the anisotropies calculated for $t_1=(0,10^3,5\times10^3,1\times10^4,2\times10^4,....,9\times10^4)yr$ at a fixed value of $t_2=10^5yr$. The fraction of the total SNE energy of $10^{51}ergs$ going into the kinetic energy of protons is taken as $f=0.1$. Data points are taken from the compilation of different experiments given in Erlykin $\&$ Wolfendale 2006. \textit{Top} : For a diffusion coefficient $D(E)\propto E^a$ with $a=0.3$. \textit{Bottom} : $a=0.6$.}
\end{figure}
\begin{figure}
\centering
\includegraphics*[width=0.31\textwidth,angle=270,clip]{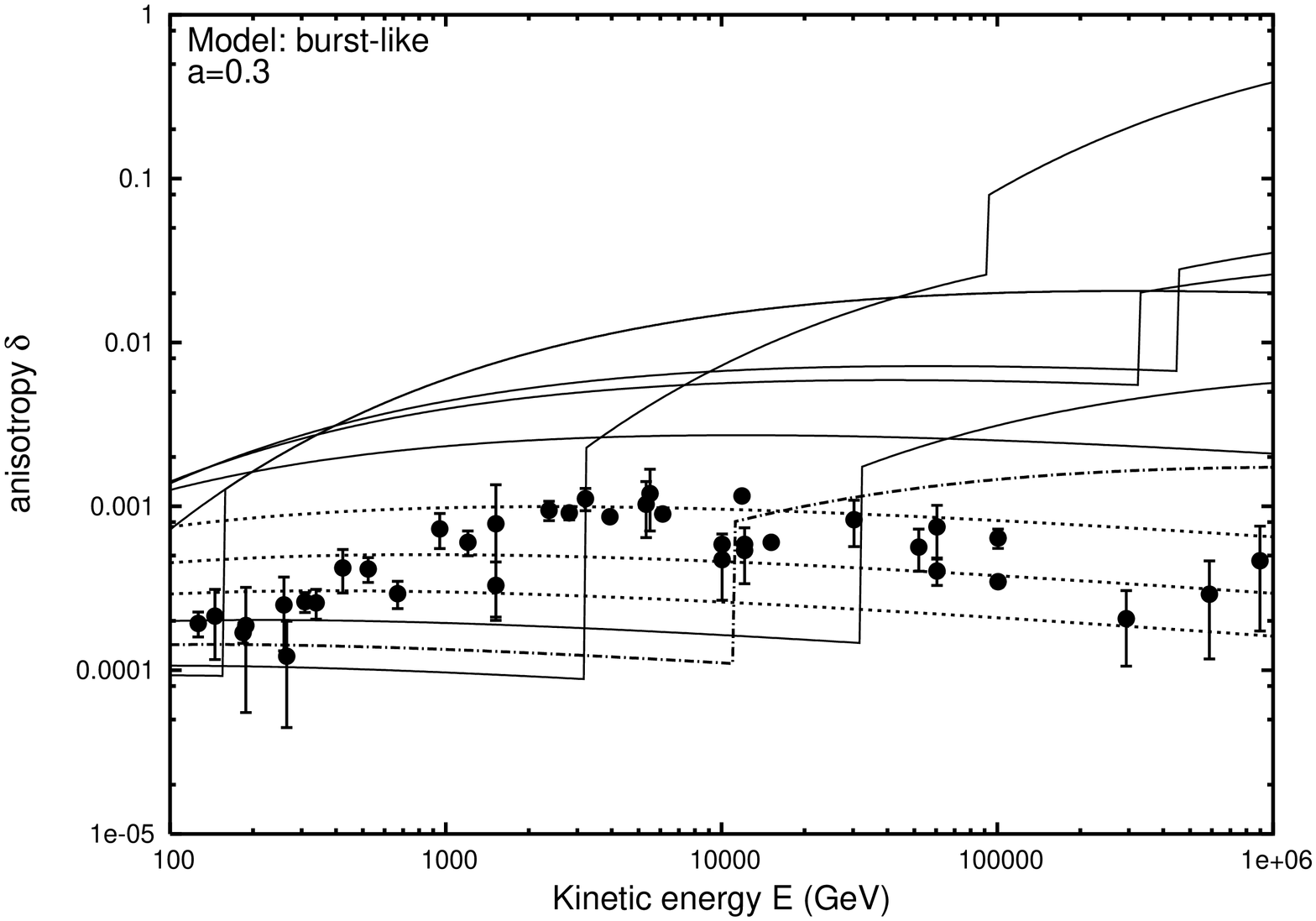}
\includegraphics*[width=0.31\textwidth,angle=270,clip]{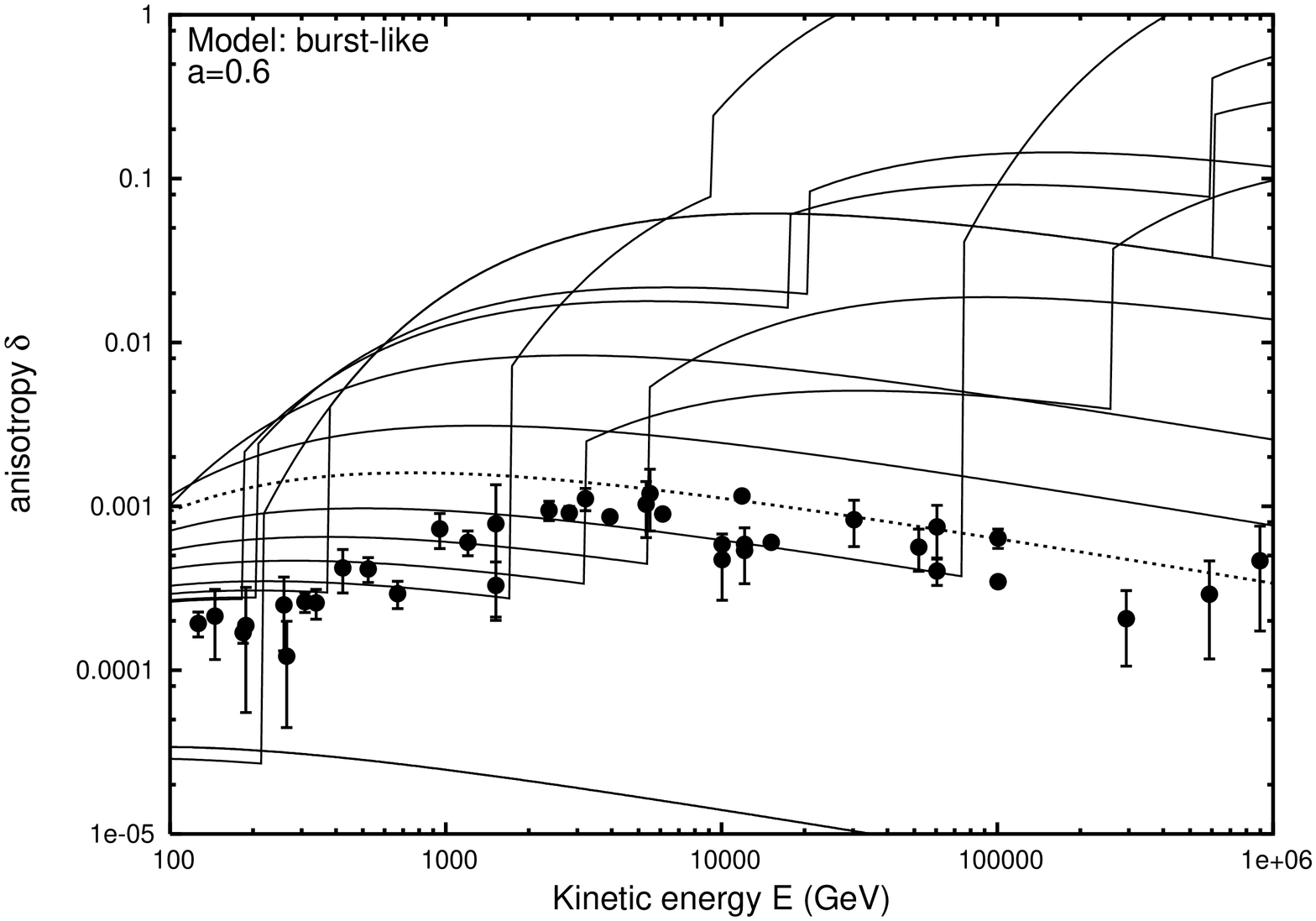}
\caption{\label {fig2} CR anisotropy $\delta$ (Eq. 8) for the burst-like injection case. The lines represent the anisotropies calculated at $t_0=(0,10^3,5\times10^3,1\times10^4,2\times10^4,....,10\times10^4)yr$. Other model parameters and the data points are the same as in Fig. 1. \textit{Top} : Diffusion coefficient $D(E)\propto E^a$ with $a=0.3$. The three dotted lines calculated at $t_0=4\times10^4yr$ (bottom-dotted line), $5\times10^4yr$ (middle-dotted line) and $6\times10^4yr$ (top-dotted line) and the dot-dashed line (calculated at $2\times10^4yr$) are those which closely explain the observed data for the acceptable range of $f=(0.1-0.3)$ (see text for details). \textit{Bottom} :  $a=0.6$. The dotted line represents the anisotropy calculated at $t_0=5\times10^4yr$ which explain the data most effectively in this case.}
\end{figure}

Figs. 1$\&$2 show the comparision of the CR anisotropy amplitude observed experimentally with the contribution of the SNRs listed in Table 1 calculated using the continuous and burst-like particle injection models, respectively. The experimental data are taken from the compilation of various results given in Erlykin $\&$ Wolfendale 2006. The model calculations use Eq. 8 and assume $D(E)\propto E^a$ with $a=0.3$ or $0.6$ and a total SNE energy of $10^{51} ergs$ with a fraction $f=0.1$ of it going into the kinetic energy of the injected protons. From Fig. 1 (bottom), it can be seen that for $a=0.6$ the model with continuous injection of particles gives an anisotropy that is too large to explain the observed value. However, for $a=0.3$ and $t_1\leq10^4yr$, the continuous injection model predicts a value which agrees with the data within a factor of 2 for energies below $10^{13}eV$ and a much larger value above it, as shown in Fig. 1 (top). However, Fig. 2 shows that the burst-like particle injection model is able to explain the data more  closely, both at $a=0.3$ and $0.6$ for some values of $t_0$. For $a=0.3$, considering the acceptable range of $f=(0.1-0.3)$, it is found that the data are explained at almost two different values of $t_0$: $t_0\sim2\times10^4yr$ and $t_0\sim 5\times10^4yr$. For $t_0=2\times10^4yr$, the total anisotropy $\delta$ is determined by the Monogem SNR for energies $E\lesssim 10^{13}eV$ and by the SNR G114.3+0.3 for $E\gtrsim 10^{13}eV$, as shown by the dot-dashed line in Fig.2 (top). This also results in a change of the direction (phase) of $\delta$ at $E\sim 10^{13}eV$. It should be noted that the position of the Monogem SNR in the Galaxy with longitude $l=201.1^\circ$ as well as that of the SNR  G114.3+0.3 $(l=114.3^\circ)$ are close to the direction of the observed phase of the anisotropy (the second quadrant of the Galaxy). However, for $t_0\sim 5\times10^4yr$, the data are explained solely by the Monogem SNR, as shown by the three dotted lines in Fig. 2 (top). For the other case (i.e. $a=0.6$), Fig. 2 (bottom) shows that the burst-like injection model with $t_0\sim 5\times10^4yr$ explains the data effectively above $10^{12}eV$ while below $10^{12}eV$ the model predicts a value that is larger than the observed value by a factor of about 4 (as shown by the dotted line). In this case, the anisotropy is also determined by the Monogem SNR, as in the case of $a=0.3$ and $t_0\sim 5\times10^4yr$ discussed above.

The figures, particularly Fig. 2, show that depending on the age and distance of the local sources, the total anisotropy can be determined by different sources at different energy ranges. These can be seen as breaks in few of the lines. Also note the similarity of the lowest lines in the bottom plots of Figs. 1 $\&$ 2. AlThough the two lines correspond to two completely different particle injection models, the similarity is because, for a continuous injection model with $t_1=9\times 10^4yr$ (and $t_2=10^5 yr$) and a burst-like injection model with $t_0=10^5 yr$, only the Geminga supernova with an estimated age of $t\sim 3.4\times 10^5 yr$ has liberated the CR particles in the local region (see Table 1). Moreover, as already mentioned, the high energy particle spectrum and the anisotropy parameter $\delta$ from Geminga in this case of continuous injection are the same as in the burst-like injection case (see Case 2 of section 3.2). 

Another probable contribution to the local CR anisotropy that we have not considered yet is that from the \textit{undetected} old SNRs. Studies assuming an adiabatic phase in the SNR evolution have shown that the surface brightness estimated from an SNR of age $\sim 10^5$ yrs lies below the detection limit of radio telescopes (Kodaira 1974, Leahy $\&$ Xinji 1989). Such old SNRs can contribute a significant flux of CR protons at the Earth, although the electrons present inside it are not energetic enough to produce a strong radio flux that can be detected by present day telescopes. Fig. 3 shows the anisotropies due to such old sources calculated by taking their ages $t=2\times10^5 yr$ (below which they are assumed to be detectable) and $a=0.3$ or $0.6$ assuming burst-like injection of particles at $t_0=10^5yr$. It is worth mentioning at this point that for such old sources the particle injection stops long before the observation time because CRs are assumed to remain confined in the sources only upto around $10^5yr$ in the present model, and hence the particle injection can be well approximated by the burst-like model (see Case 2 of section 3.2). The anisotropy at each energy point in Fig. 3 is calculated using Eq. 8 and by taking the source position at a distance $r=r_{max}(E)$, where $r_{max}(E)=\sqrt{6D(t-t_0)}$ is the source distance giving the maximum flux of CRs of energy $E$ at the Earth for a given $t$ and $t_0$. Because $\delta$ for this case follows $\delta\propto(t-t_0)^{-2}$, its value increases or decreases with $t_0(t)$ for a fixed $t(t_0)$. For $t_0=10^5yr$  the maximum CR source confinement time and $t=2\times10^5yr$, the lowest possible age of the \textit{undetected} old SNRs adopted in the present study, each point in the plot represents the maximum possible anisotropy amplitude at energy $E$ due to the old sources in the present model. Calculations using other possible values of the source parameters (i.e. $t_0<10^5yr$ and $t>2\times10^5yr$) will give anisotropies that are less than the values shown in Fig. 3. In Fig. 3, the solid lines represent the model with $a=0.3$ and the dotted lines are for $a=0.6$; the upper lines represent the values calculated at $f=0.3$ and the lower lines at $f=0.1$. The observed anisotropy data are also shown in order to see the extent of the contribution of the old sources. A detailed inspection of the figure shows that the contribution of the \textit{undetected} old SNRs to the CR anisotropy at the Earth cannot be fully neglected, particularly for the model with $a=0.6$. This can be seen from the region bounded by the two dotted lines, which contains the data points below $E\lesssim 10^{12}eV$ and above $E\gtrsim 10^{14}eV$. However, here it should be mentioned again that each point given by the calculated lines only represents the maximum possible anisotropy due to the old sources. Physically, it is highly unrealistic to get such a situation at any point of time, and thereby each line represents the values observed at the Earth in the whole energy range. However, there is always a possibility of observing atleast one of the energy points at a given time. Thus, we cannot claim that the \textit{undetected} old sources can fully explain the observed anisotropy data, but we can safely conclude that their contribution cannot be completely neglected in the study of CR anisotropy at the Earth.  

There also exists a model dependent argument, which supports the contribution of the old sources of the observed CRs. For the generally excepted particle release time of $t_0=10^5yr$ (which is also the value adopted here), all the nearby known SNRs listed in Table 1 are quite young (with estimated ages less than $10^5 yr$) except for the Geminga SNR. Thus, they have not yet released the CRs into the local ISM. So, it is quite possible that majority of the CRs observed at the Earth originate from nearby \textit{undetected} old sources. In fact, the necessity for the presence of such old nearby sources has already been disccussed in the spectral study of both CR protons (Thoudam 2006) as well as CR electrons (Atoyan et al. 1995). 

\begin{figure}
\centering
\includegraphics*[width=0.31\textwidth,angle=270,clip]{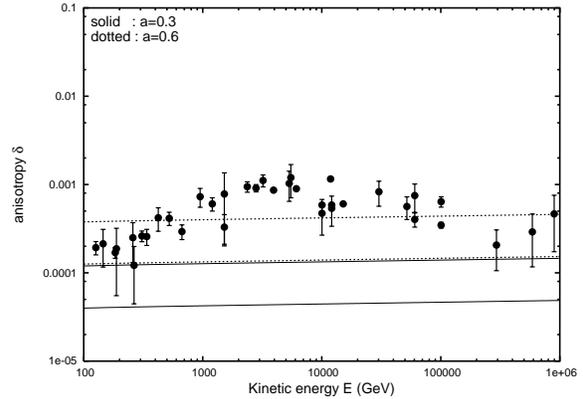}
\caption{\label {fig3} Maximum CR anisotropy at the Earth due to \textit{undetected} old SNRs with ages $t=2\times10^5yr$ assuming burst-like injection of particles at $t_0=10^5yr$. The calculations are done using Eq. 8 and  by taking the source distances $r=r_{max}(E)$, where $r_{max}(E)=\sqrt{6D(t-t_0)}$ is the source distance which gives the maximum flux of CRs of energy $E$ at the Earth for a given $t$ and $t_0$. Each point represents the maximum possible anisotropy $\delta$ from an old source at a given energy $E$ (see text for details). The solid lines are for $a=0.3$ and the dotted lines for $a=0.6$ in which the upper ones represent the values at $f=0.3$ and the lower ones at $f=0.1$. Data points are the same as in Fig. 1. and are shown just to see the extend of the old sources contribution.}
\end{figure}

\section{Results and discussions}
The analysis presented here involves the study of the effect of different particle release times from the sources on the CR spectrum and the anisotropy below $\sim 10^{15}eV$ observed at the Earth. The study considers burst-like injection, continuous injection for a finite time, injection from a stationary source and energy dependent injection as possible forms of particle injection from the sources. The study shows that the case of particle injection from a stationary source is the same as that of the continuous injection for a finite time period for which $t_1<t\leq t_2$ (see Case 1 of section 3.2). Also, for the proton energy range of $(0-10^{15})eV$ considered here, the energy dependent particle injection model can be represented by the burst-like injection of particles at time $t_0=10^5yr$ except for particles with energies $E\geq 4\times 10^{14}eV$. When applied to the nearby known SNRs located within a distance of $1.5 kpc$, it is found that the model with continuous injection of particles gives an anisotropy that is too large to explain the observed data at $a=0.6$, but it predicts a value that is within a factor of 2 for $E\leq 10^{13}eV$ at $a=0.3$ and $t_1\leq10^4yr$. However, the burst-like injection model with $a=0.3$ can explain the data more closely at almost two values of particle injection time: $t_0\sim 2\times10^4yr$, where the anisotropy is determined by Monogem for $E\lesssim 10^{13}eV$ and by the source G114.3+0.3 for $E\gtrsim 10^{13}eV$ and $t_0\sim 5\times10^4yr$, where Monogem dominates in the whole energy range below $10^{15}eV$. Also, the burst-like model with $a=0.6$ explains the data effectively above $10^{12}eV$ if particle injection takes place at $\sim 5\times10^4yr$, although below $10^{12}eV$ it predicts an anisotropy that is larger than the observed value by a factor of almost 4. For this case, the total anisotropy is again determined mainly by the Monogem SNR. This result is strongly supported by the close agreement of the observed phase of the anisotropy with the positions of G114.3+0.3 and Monogem in the Galaxy. Therefore, if the observed anisotropy is due to the effect of the local sources, the study favours the burst-like particle injection model both at $a=0.3$ and $0.6$. 

The study further shows that the contribution of the \textit{undetected} old SNRs cannot be fully neglected in the study of CR anisotropy at the Earth. Because of their unknown source parameters (such as their ages and distances) it is impossible to calculate their actual contribution to the observed CRs. However, we can always try to make an estimate of their contribution by considering a proper model dependent set of source parameters. In the present study, using source ages of $t=2\times10^5yr$ and particle release time of $t_0=10^5yr$, it has been shown that for the model with $a=0.6$ they can contribute significantly to the local CRs. Calculations using other possible values of the source parameters (i.e. $t_0<10^5yr$ and $t>2\times10^5yr$) will give anisotropies that are less than the values obtained here. Thus, Fig. 3 shows only the maximum possible anisotropy value at energy $E$ due to the old sources in the present model. The study concludes that the contribution of the \textit{undetected} old SNRs to the observed CR anisotropy cannot be simply ignored and should be considered in the study of local CRs. 

\section{Acknowledgements}
The author is very grateful to an anonymous referee for helpful suggestions.

\end{document}